\documentclass[aps,twocolumn,superscriptaddress,floatfix,letterpaper]{revtex4-1}

\usepackage{amsmath, amssymb, braket}

\newcommand{\RZ}{{\mathbb{R}/\mathbb{Z}}}

\newcommand\hcup[1]{\underset{{\scriptscriptstyle #1}}{\smile}}
\newcommand\toZ[1]{\lfloor #1 \rceil}

\newcommand\Ref[1]{Ref. \cite{#1}}

\newcommand\beq{\begin{equation}}
\newcommand\eeq{\end{equation}}

\usepackage[dvipsnames]{xcolor}

\newcommand{\ie}{\emph{i.e.} }
\newcommand\eqn[1]{eqn. \ref{#1} }
\newcommand\eq[1]{eqn. \ref{#1} }
\newcommand{\dd}{\text{d}}
\newcommand\cM{\mathcal{M}}
\newcommand\etc{\emph{etc.}}
\newcommand\ee{\text{e}}
\newcommand\ii{\text{i}}
\newcommand\R{\mathbb{R}}
\newcommand\Z{\mathbb{Z}}

\begin{document}

\begin{titlepage}

\title{
Lattice realization of compact $U(1)$ Chern-Simons theory
with exact 1-symmetries
}

\author{Michael DeMarco}
\affiliation{Department of Physics, Massachusetts Institute of
Technology, Cambridge, MA 02139, USA}
\email{demarco@mit.edu}

\author{Xiao-Gang Wen}
\affiliation{Department of Physics, Massachusetts Institute of
Technology, Cambridge, MA 02139, USA}
\email{xgwen@mit.edu}

\begin{abstract} 

We propose a bosonic $U(1)$ rotor model on a three dimensional space-time
lattice.  With the inclusion of a Maxwell term, we show that the low-energy
semi-classical behavior of our model is consistent with the Chern-Simons field
theory, $S= \int \text{d}^3 x\ \frac{K_{IJ}}{4\pi} A_I \text{d} A_J$ at low
energies. We require that the action be periodic in the lattice variables,
which enforces the quantization of the $K$-matrix as a symmetric integer matrix
with even diagonals. We also show that our lattice model has the exact
$1$-symmetries.  In particular, some of those 1-symmetries are anomalous (\ie
non-on-site) in the expected way.  The anomaly can be probed via the breaking
of those $1$-symmetries by the boundaries of space-time.

\end{abstract}


\maketitle

\end{titlepage}


\noindent
\textbf{Introduction}:
Chern-Simons (CS) field theory is a very important field theory with
myriad applications from condensed matter to quantum gravity. Though well
studied in the continuum as field theory, defining CS field theory on the
lattice presents an opportunity to better tame the field integration measure
as well as allowing us to consider non-smooth gauge fields with singularities.
Furthermore, it is well known that it is quite non-trivial to define the action of CS
field theory if the fiber bundle described by the gauge field is non-trivial on
the space-time,\cite{DW9093} which leads to an obstruction to have a globally
defined gauge fields (the connection 1-forms). One way to address this problem
is to define CS theory on a space-time lattice where the lattice gauge fields for
``distinct fiber bundles'' are continuously connected.  Then the lattice gauge
field with monopoles/flux-lines is continuously connected to gauge field
without monopoles/flux-lines.  So once we have a space-time lattice description
of CS theory, the theory is automatically well defined for gauge fields of
topologically non-trivial fiber bundle, as well as for gauge fields of
monopoles and flux-lines.  Certainly, the lattice description of CS field
theory also remove the infinity problem of the field theory.  

One natural way to put CS theory on spatial lattice with continuous time is to
consider lattice boson or fermion systems. If the bosons or fermions are in a
quantum Hall state, then the topological order of the  quantum Hall state is
described by a low energy effective  CS field theory of a compact gauge group.
However, the  boson or fermion lattice models are usually not soluble.  Given a
boson or fermion lattice model, we usually do not know if it is in a
quantum-Hall topologically ordered phase.  We usually do not know if the
lattice model produce a CS theory at low energy or not, and we do not know
which CS theory it produces. So here we are looking for a better result, where
we can derive, under a controlled approximation, the low energy effective CS
field theory from the lattice model.

A related lattice formulation of the CS action has been proposed in
\Ref{ESh9204048} for spatial lattice and continuous time.  The proposed lattice
model is not a local bosonic model, but a lattice gauge theory with a Hilbert
space that does not have a tensor product decomposition (\ie the total
many-body Hilbert space of the lattice model is not a tensor product of local
Hilbert space on each site).  This is because the construction in
\Ref{ESh9204048} impose the $U(1)$ gauge equivalence generated by the ``small''
and ``large'' $U(1)$ gauge transformations, and thus the path integration is
not the product of integrals over local variables (and hence the non-locality
of the lattice model). The constructed (non-local) lattice model describes a
CS theory of a compact $U(1)$ on a space with no boundary.
Furthermore, since the gauge invariance is broken on space with boundary, it
remains to figure out how to define a CS theory of compact $U(1)$ on
a space with boundary.  In comparison, this paper presents a different approach
by starting from a \emph{local} space-time lattice model (\ie the path
integration is the product of integrals over independent local variables),
which also allows us to define a CS theory of compact $U(1)$ on a space with
boundary.  In short, the model constructed in \Ref{ESh9204048} is a lattice
gauge theory which is non-local, while the model constructed in this paper is a
local bosonic model on space-time lattice.


\Ref{BSh0004203} tried to put CS theory of \emph{non-compact} $U(1)$
on space-time lattice as a quadratic (non-interacting) theory.  
Another lattice formulations of the CS action have been proposed in
\Ref{BSl0207010}  for space-time lattice.  Such a theory also has a non-compact
$U(1)$. Furthermore, as a lattice gauge theory, the space-time lattice model is
not local. For a CS theory with non-compact $U(1)$, its ground state
degeneracy (as a quantized theory) on a torus will be infinity, suggesting that
the lattice model, as a quantum theory, is not well defined since no
well-defined local lattice quantum theory can produce non-compact $U(1)$
CS theory as low energy effective field theory.

We see that people tried to put CS theory on lattice for almost 30
years. Despite those efforts, putting an arbitrary compact-$U(1)$ CS
theory on space-time lattice remain an unsolved problem, if we require that
that we can \emph{reliably} determine the low energy effective compact $U(1)$
CS field theory from a \emph{local} lattice model (\ie a model where
the field integral is a product of integrals of local variables).  In this
paper, we will propose a solution to this problem.

We will propose a well-defined local bosonic model on space-time lattice
\eqn{CSlatt}.  Under a controlled semi-classical approximation for small $g$
in \eqn{CSlatt}, we show that our  space-time lattice model can produce any
even-$K$-matrix CS field theory\cite{WZ9290} of compact $U(1)$'s in
continuum limit (see \eqn{CSKIJ}).  We will relies on the cochain theory
familiar from algebraic topology \cite{Ha02} to construct our lattice model.
While formulating ``$a\dd a$'' on the lattice is straightforward, the key to
this approach is insuring that the lattice variables remain periodic and that
the action is invariant under the period (\ie the $U(1)$'s are compact), even
for space-time with boundary.

A striking character of lattice model \eq{CSlatt} is that the Lagrangian
density is not a continuous function of the field values.  Also the lattice
model is defined for $\RZ$-valued fields and is not quadratic (\ie corresponds
to an interacting theory).  But the weak fluctuations in small $g$ limit are
described by a quadratic action.  This is why we can reliably obtain the low
energy effective CS field theory in  small $g$ limit.

However, being able to reliably obtain the low energy effective theory is not
the most important character of our constructed model \eq{CSlatt}.  What really
special of our lattice model is that it has many exact 1-symmetries.  It was
known that 2+1D $Z_n$ gauge theory described by mutual CS theory has many
1-symmetries on
lattice.\cite{K032,W0303,LW0316,HW0541,Y10074601,B11072707,W181202517} A
generic $U(1)$ CS theory also has many 1-form symmetries in
continuum.\cite{GW14125148,HS181204716} Our lattice realization of a generic
$U(1)$ CS theory is a very special one, that those 1-form symmetries in
continuum become the exact 1-symmetries\cite{W181202517} in our lattice model.
In contrast, the quantum Hall realization of CS theory does not have those
1-symmetries.  Therefore, we can state our result more precisely as the
following:
\begin{center}
\framebox{\parbox{.97\columnwidth}{We construct a local bosonic rotor model on space-time
lattice that, at low energies, realizes most general compact $U(1)$ CS field
theory characterized by $K$-matrix, where the 1-form symmetries in the CS field
theory are realized as the exact $Z_{k_1}\times Z_{k_2} \times \cdots$
1-symmetries in our lattice model.\footnote{There is a small difference between
1-form symmetry\cite{GW14125148} and 1-symmetry\cite{W181202517}
transformations, although they both act on closed sub-manifold with
codimension-1. The 1-form symmetry transformation acting contractable
sub-manifold is a trivial transformation (\ie an identity operator), while the
1-symmetry transformation acting contractable sub-manifold is a non-trivial
transformation not equal to identity operator.} Here $k_i$ are diagonal entries
of the Smith normal form of the $K$-matrix.  Some of the 1-symmetries are
anomalous.\cite{TK151102929,BM170200868,KR180505367,W181202517}  Different
lattice realizations of the same $U(1)$ CS field theory may lead to diiferent
anomalous 1-symmetries.}}
\end{center}

We like to remark that our model is not a gauge theory, since it is defined via
a path integral that is a product of integrals of local variables.  We build
our model to be periodic in the lattice variables, and take the field integral
over one period of each lattice variable. This periodic redundancy provides
level quantization, without relying on the underlying topology of the manifold
as ``large gauge transformations'' do.  Since our lattice model is not a
lattice gauge theory, its action does not has to be gauge invariant.  Indeed,
the action \eqn{CSlatt}, on space-time with boundary, is not invariant under
the usual gauge redundancy $a\to a + \dd \theta$, with $\theta$ a $0$-cochain.

We also like to remark that our lattice model \eqn{CSlatt} is actually a
tensor network path integral in space-time.\cite{GW0931}  Thus, we have found a
tensor network path integral that realize a topologically ordered phase
described by CS field theory.  We note that the tensor in the tensor network is
indexed by a $\RZ$-value.  In other words, the dimension of the tensor is
infinity.

%

\noindent
\textbf{Chern Simons Theory on Lattice}:
To construct our local bosonic space-time lattice model, we will use a cochain
formalism on a space-time complex.  A space-time complex (lattice)  is a
triangulation of the three-dimensional space-time \emph{with a branching
structure},\cite{C0527,CGL1314,CGL1204} which is denoted as $\cM^3$.  The
space-time complex is formed by simplices -- the vertices, links, triangles,
\etc.  We will use $i,j,\cdots$ to label vertices of the space-time complex.
The links of the complex (the 1-simplices) will be labeled by $\braket{ij},\cdots$.
Similarly, the triangles of the complex (the 2-simplices) will be labeled by
$\braket{ijk},\cdots$.  The degrees of freedom of lattice model live on the links of
the space-time complex: $(a^{\RZ}_I)_{ij}$ on link $\braket{ij}$,
$I=1,2,\cdots,\kappa$.  $(a^{\RZ}_I)_{ij}$ is $\RZ$-valued, \ie $(a^{\RZ}_I)_{ij}$
and $(a^{\RZ}_I)_{ij}$ are equivalent if $(
a^{\RZ}_I)_{ij}-(a^{\RZ}_I)_{ij}=0$ mod 1.  Such $\RZ$-valued fields on the
links are simply the so called 1-cohains $a^{\RZ}_I$ on the space-time
complex $\cM^3$.  Here we have $\kappa$ different 1-cochains $a^{\RZ}_I$ labeled
by $I$. The lattice action of our bosonic model will be constructed from those
1-cochains using cup product and cochain derivative.  For a more detailed
introduction to the cochain formalism for defining local bosonic space-time
lattice models, see \Ref{W161201418} and Appendix \ref{sec:CaC}.  

We want to construct our lattice bosonic model in such a way that it is very
similar to a CS theory.  Hopefully, the resulting lattice bosonic
model realizes a topologically ordered state described the CS
topological quantum field theory.  Due to the similarity between 1-cochain and
differential 1-form, between the cup product for cochains and wedge product for
differential forms, as well as the derivative $\dd$ acting on them,
na\"{i}vely, we would write the partition function for a bosonic lattice as:
\begin{align}
Z=\int [\prod\dd a^{\RZ}_I]
\ee^{\ii 2\pi \sum_{I\leq J} k_{IJ} \int_{\cM^3} a^{\RZ}_{I} \dd a^{\RZ}_{J}},
\label{eq:NaiveCS}
\end{align}
which formally looks like the continuum CS field theory written in
terms of differential 1-forms.  Here $k_{IJ}$ are integers, and $\int_{\cM^3}$
means the sum over all $3$-simplices in $\cM^3$.  Also $\int [\prod\dd a^{\RZ}_I]
\equiv \prod_{\braket{ij}}\prod_I \int_{-\frac12}^{\frac12}\dd (a^{\RZ}_I)_{ij}$
gives rise to the path integral, where $\prod_{\braket{ij}}$ is a product over all
the links.  Here we have lifted the $\R/\Z$-valued $(a^{\RZ}_I)_{ij}$ to a
$\R$-valued $(a^{\RZ}_I)_{ij}\in (-\frac12,\frac12]$ before we do the path
integral.

Since $a^{\RZ}_I$ is $\R/\Z$-valued, we require the action amplitude in
\eqn{eq:NaiveCS} to be invariant under the following ``gauge'' transformation
\begin{align}
\label{gaugeZ}
 a^{\RZ}_I \to a^{\RZ}_I + n^I,
\end{align}
where $n^I$ are arbitrary $\Z$-valued 1-cochains.  But, the action amplitude
$\ee^{\ii 2\pi \sum_{I\leq J} k_{IJ} \int_{\cM^3} a^{\RZ}_{I} \dd a^{\RZ}_{J}}$
is not gauge invariant and we need to fix it. For a bosonic system, with
$k_{IJ}\in \mathbb{Z}$, one way to fix this problem is to consider the following
modified partition function (which is the main result of this paper):
\begin{align}
\label{CSlatt}
& Z =\int [\prod\dd a^{\RZ}_I]\ 
\ee^{\ii 2\pi \sum_{I\leq J} k_{IJ} \int_{\cM^3} \dd \big(a^{\RZ}_I(a^{\RZ}_J-\toZ{a^{\RZ}_J}  )\big) }
\nonumber\\
&
 \ee^{\ii 2\pi \sum_{I\leq J} k_{IJ} \int_{\cM^3} a^{\RZ}_{I} (\dd a^{\RZ}_{J} -\toZ{\dd a^{\RZ}_J})-\toZ{\dd a^{\RZ}_I}a^{\RZ}_J }
\\
&
\ee^{-\ii 2\pi \sum_{I\leq J} k_{IJ} \int_{\cM^3} a^{\RZ}_J\hcup{1}\dd \toZ{\dd a^{\RZ}_I}}  
\ee^{- \int_{\cM^3} \frac{|\dd a^{\RZ}_I - \toZ{\dd a^{\RZ}_I}|^2}{g}}
\nonumber 
\end{align}
Here $\toZ{x}$ denotes the nearest integer to $x$ and $\toZ{\dd a^{\RZ}_I}$ is
the 2-cochain whose value on the triangle $\braket{ijk}$ is given by $\toZ{(\dd
a^{\RZ}_I)_{ijk}}$.  The 1-cup product $\hcup{1}$ is defined in Appendix
\ref{sec:CaC}.\cite{S4790}  We note that when $\dd a^{\RZ}_I \approx 0$,
\eqn{CSlatt} reduces to \eqn{eq:NaiveCS}.  The Maxwell term $\ee^{-
\int_{\cM^3} \frac{|\dd a^{\RZ}_I - \toZ{\dd a^{\RZ}_I}|^2}{g}}$ is included to
make $\dd a^{\RZ}_I$ nearly an integer if we choose $g$ to be small. 

To see that the path integral \eq{CSlatt} is invariant under gauge
transformation \eq{gaugeZ} for $\cM^3$ with boundary, we first note that
$\ee^{-\ii 2\pi \sum_{I\leq J} k_{IJ} \int_{\cM^3} a^{\RZ}_J\hcup{1}\dd
\toZ{\dd a^{\RZ}_I}}$ and $\dd a^{\RZ}_I - \toZ{\dd a^{\RZ}_I}$ are invariant
under \eqn{gaugeZ}.  Under \eqn{gaugeZ}, the term $\ee^{\ii 2\pi \sum_{I\leq J}
k_{IJ} \int_{\cM^3} a^{\RZ}_{I} (\dd a^{\RZ}_{J} -\toZ{\dd a^{\RZ}_J})-\toZ{\dd
a^{\RZ}_I}a^{\RZ}_J } $ changes by a factor
\begin{align}
&\ \ \ \
\ee^{\ii 2\pi \sum_{I\leq J} k_{IJ} \int_{\cM^3} n^{I} \dd a^{\RZ}_{J} -\dd n^Ia^{\RZ}_J }
\nonumber\\
&=
\ee^{-\ii 2\pi \sum_{I\leq J} k_{IJ} \int_{\partial\cM^3} n^Ia^{\RZ}_J }
\end{align}
Such a factor is cancelled by the change of the term 
\begin{align}
&\ \ \ \
 \ee^{\ii 2\pi \sum_{I\leq J} k_{IJ} \int_{\cM^3} \dd \big(a^{\RZ}_I(a^{\RZ}_J-\toZ{a^{\RZ}_J})\big) } 
\nonumber\\
&
=
 \ee^{\ii 2\pi \sum_{I\leq J} k_{IJ} \int_{\partial \cM^3} \big(a^{\RZ}_I(a^{\RZ}_J-\toZ{a^{\RZ}_J})\big) } .
\end{align}
So the action amplitude of the above path integral is indeed invariant under $\eq{gaugeZ}$ even when $\cM^3$ has 
boundary.

Now, we like to argue that the bosonic lattice model \eq{CSlatt} realizes a
topological order described by $U^\kappa(1)$ CS topological quantum
field theory, in the small $g$ limit.  In such a limit, $\dd a^{\RZ}_I$ is
close to an $\Z$-valued cocycle.  On a local patch of space-time, we use the
gauge transformation \eqn{gaugeZ} to make $\dd a^{\RZ}_I$ to be near zero on
the patch.  In this case, the action amplitude in the path integral
\eqn{CSlatt} becomes quadratic (\ie non-interacting) 
\begin{align}
\ee^{\ii 2\pi \sum_{I\leq J} k_{IJ} \int_{\cM^3} a^{\RZ}_{I} \dd a^{\RZ}_{J}}.
\end{align}
Since $\dd a^{\RZ}_I$ is close to zero,
we can use a 1-form $A^I$ to describe the 1-cochain $a^{\RZ}_I$:
\begin{align}
 \int_i^j A^I = 2\pi(a^{\RZ}_I)_{ij}
\end{align}
Then the above action amplitude can be rewritten as
\begin{align}
\label{CSKIJ}
&
\ee^{\ii 2\pi \sum_{I\leq J} k_{IJ} \int_{\cM^3} a^{\RZ}_{I} \dd a^{\RZ}_{J}}
\approx \ee^{\ii \sum_{I J} \frac{K_{IJ}}{4\pi} \int_{M^3} A^{I} \dd A^{J}}
\nonumber\\
&\ \ \ \
K_{IJ} = K_{JI} \equiv \begin{cases}
2 k_{IJ}, & \text{ if } I=J,\\
k_{IJ}, & \text{ if } I<J,\\
\end{cases}
\end{align}
in the small $\dd a^{\RZ}_I$ limit when $A^I$ is nearly constant on the lattice
scale. Hence the low energy dynamics of our lattice bosonic model are described
by a $U^\kappa(1)$ CS field theory \eq{CSKIJ} at low energies.

We like  to remark that, when $\dd a^{\RZ}_I$ is near integers, $\toZ{\dd
a^{\RZ}_I}$ is a $\Z$-valued 2-cocycle.  This is because if $\dd a^{\RZ}_{I} =
\epsilon +\toZ{\dd a^{\RZ}_{I}}$ where $\epsilon$ is small, then
\begin{align}
 \dd \toZ{\dd a^{\RZ}_{I}} = -\dd \epsilon +\dd \dd a^{\RZ}_I =-\dd \epsilon.
\end{align}
Since $\dd \toZ{\dd a^{\RZ}_{I}}$ is quantized as integer, we have
\begin{align}
 \dd \toZ{\dd a^{\RZ}_{I}}=0.
\end{align}
Such a $\Z$-valued 2-cocycle $\toZ{\dd
a^{\RZ}_I}$ characterize the $U^\kappa(1)$
principle bundle on the space-time, since
\begin{align}
 \int_{\cM^2} (\dd a^{\RZ}_I-\toZ{\dd a^{\RZ}_I})=
 -\int_{\cM^2} \toZ{\dd a^{\RZ}_I}
\end{align}
for any closed $\cM^2$.  Note that $\int_{\cM^2} (\dd a^{\RZ}_I-\toZ{\dd
a^{\RZ}_I})$ is the magnetic flux through $\cM^2$ which is always quantized to be an integer. In other words, $-\int_{\cM^2} \toZ{\dd a^{\RZ}_I}$ is the Chern number.

The above discussion of dynamics only apply when $\dd a^{\RZ}_I$ is near
integers, \ie when $g$ is small.  When $g$ is large, the large quantum
fluctuations of $a^{\RZ}_I$ in the lattice bosonic model can go between
configurations representing different $U^\kappa(1)$ principle bundles.  The large
$g$ ground state of our model \eq{CSlatt} may have a different topological
order from the one described by the $K$-matrix CS theory.

What is really special about our constructed action is that it has many
1-symmetries. First, consider the model on a closed manifold, so that we may
ignore the surface term. Then under the shift
\begin{align}
\label{1symm}
a^{\RZ}_{I} \to a^{\RZ}_{I} + \beta^{\RZ}_{I},\ \ \ \sum_I \beta^{\RZ}_{I}K_{IJ} \in \Z
\end{align}
where $\beta^{\RZ}_{I}$ are $\RZ$-valued 1-cocycles, the exponentiated action
is invariant so long as $\sum_I \beta^{\RZ}_{I}K_{IJ}$ are $\Z$-valued
1-cochain.  Such the transformations \eq{1symm} are the 1-symmetries of
lattice model \eq{CSlatt}.  

To see the above result, we first note that, under the transformation
\eq{1symm}, the action amplitude in \eqn{CSlatt} on a closed manifold changes
by a factor
\begin{align}
& \ee^{\ii 2\pi \sum_{I\leq J} k_{IJ} \int_{\cM^3} \beta^{\RZ}_{I} (\dd a^{\RZ}_{J} -\toZ{\dd a^{\RZ}_J})-\toZ{\dd a^{\RZ}_I}\beta^{\RZ}_J} \times
\nonumber\\
& \ee^{-\ii 2\pi \sum_{I\leq J} k_{IJ} \int_{\cM^3} \beta^{\RZ}_J\hcup{1}\dd \toZ{\dd a^{\RZ}_I}} 
\end{align}
Because we may integrate by parts on a closed manifold and $\dd\beta^\RZ_{I} = 0$,
the change is of the form (see \eqn{cupkrel}):
\begin{align}
& \ee^{-\ii 2\pi \sum_{I\leq J} k_{IJ} \int_{\cM^3} \beta^\RZ_{I} \toZ{\dd a^{\RZ}_J} + \toZ{\dd a^{\RZ}_I}\beta^\RZ_J +\beta^{\RZ}_J\hcup{1}\dd \toZ{\dd a^{\RZ}_I}}
\nonumber\\
&=  \ee^{-\ii 2\pi \sum_{IJ} K_{IJ} \int_{\cM^3} \beta^\RZ_{I} \toZ{\dd a^{\RZ}_J} }
\end{align}
which remains unity for all $\toZ{\dd a^{\RZ}_{J}}$ iff $\sum_I
\beta^{\RZ}_{I}K_{IJ}$ are $\Z$-valued cochains.  We see that, on a fixed link
$ij$, the allowed values $(\beta^{\RZ}_{I})_{ij}$ form the rational lattice
$K^{-1}$.  The 1-symmetries are given by the rational lattice $K^{-1}$ mod out
the integer lattice, which is same as integer lattice mod out lattice $K$.  In
other words, the 1-symmetries are $Z_{k_1}\times Z_{k_2} \times \cdots$
1-symmetries with $k_i$ being the diagonal entries of the Smith normal form of
$K$.

For example, for $U(1)$ Chern Simons theory with $\kappa = 1$ and $K_{11}
=2k_{11} =k$, we have a $\mathbb{Z}_{k}$ 1-symmetry. For mutual CS theory (that
describes a $Z_n$ gauge theory), with 
$(K_{IJ}) = \left(\begin{array}{cc}0 & n\\ n & 0\end{array}\right)$
we have a ${Z}_{n}\times {Z}_{n}$ 1-symmetry. 

Some of the above 1-symmetries are anomalous.  To see which 1-symmetries are
anomalous, we need check which of the transformations in \eqn{1symm} changes
the action amplitude when the space-time has a boundary.
Under the transformation \eq{1symm}, the action amplitude in \eqn{CSlatt} only
changes by a factor defined on the boundary $\partial \cM^3$:
\begin{widetext}
\begin{align}
&\ \ \ \
\ee^{\ii 2\pi \sum_{I\leq J} k_{IJ} \int_{\partial \cM^3} 
a^{\RZ}_I(\beta^{\RZ}_J-\toZ{\beta^{\RZ}_J})
+\beta^{\RZ}_I(a^{\RZ}_J-\toZ{a^{\RZ}_J})
+\beta^{\RZ}_I(\beta^{\RZ}_J-\toZ{\beta^{\RZ}_J})
}
\ee^{\ii 2\pi \sum_{I\leq J} k_{IJ} \int_{\partial \cM^3} 
\beta^\RZ_J \hcup{1} \toZ{\dd a^{\RZ}_I}-\beta^\RZ_I a^\RZ_J
}
\nonumber\\
&=
\ee^{\ii 2\pi \sum_{I\leq J} k_{IJ} \int_{\partial \cM^3} 
a^{\RZ}_I(\beta^{\RZ}_J-\toZ{\beta^{\RZ}_J})
+\beta^{\RZ}_I(\beta^{\RZ}_J-\toZ{\beta^{\RZ}_J})
+\beta^\RZ_J \hcup{1} \toZ{\dd a^{\RZ}_I}
}
\ee^{-\ii 2\pi \sum_{I\leq J} k_{IJ} \int_{\partial \cM^3} 
\beta^{\RZ}_I\toZ{a^{\RZ}_J}
}
\end{align}
\end{widetext}
We see that the transformations leave the action amplitude invariant if
$\sum_{I\leq J} k_{IJ} \beta^\RZ_J = 0,\ \sum_{I\leq J}
k_{IJ} \beta^\RZ_I = \text{integer}$.  We note that $\beta^\RZ_I$ satisfy the
condition $\sum_{I J} K_{IJ} \beta^\RZ_I = \text{integer} $. Thus the first
equation implies the second one.  We find that the 1-symmetry transformations in
\eqn{1symm} are anomaly-free if
\begin{align}
 \label{AFc}
 \sum_{I\leq J, L} k_{IJ} \beta^\RZ_J = 0
\end{align}

For the level $k=K_{11}$ CS theory with a single $U(1)$ gauge field, this is
simply the fact that the only $Z_k$ 1-symmetry must break at the boundary and
is anomalous.  For the case of mutual CS theory (ie the $Z_n$ gauge theory)
with $Z_n\times Z_n$ 1-symmetry, this implies that one of the $Z_n$ 1-symmetry
must break at the boundary and is anomalous.  The other $Z_n$ 1-symmetry
is anomaly-free.  Note that the choice of lattice model automatically selects
which of the $Z_n$ 1-symmetry is anomalous; one can select the opposite by
replacing all $\sum_{I\leq J}$ with $\sum_{I\geq J}$.  

\noindent
\textbf{Strong coupling limit}:
We have argued that the weak coupling limit ($g\to 0$) of our model \eq{CSlatt}
gives rise to a topological order described by Abelian CS field theory
\eq{CSKIJ}.  For invertible $K$-matrix, our model always has exact anomalous
1-symmetry on lattice.  Therefore, in the strong coupling limit $g\to \infty$,
our model is either in a phase where the anomalous 1-symmetries are
spontaneously broken,\cite{W181202517} or a gapless phase. Symmtric trivial
product state cannot be the ground state for our model.  Since spontaneously
broken higher symmetry is nothing but topological order,\cite{W181202517} in
the strong coupling limit $g\to \infty$, our model must be in a topologically
ordered phase or a gapless phase.

\noindent
\textbf{Framing anomaly}:
It is well known that the CS theory has a framing
anomaly.\cite{W8951,GF14106812} In other words, after integrating out the
physical degrees of freedom $a^\RZ_I$ in \eqn{CSlatt} in small $g$ limit, we
should get a partition function given by the 2+1D gravitational CS term:
\begin{align}
 Z(M^3,g_{\mu\nu}) \propto
\ee^{\ii \frac{2\pi c}{24} \int_{M^3} \omega_3} 
\end{align}
where the 3-form $\omega_3$ satisfies $\dd \omega_3 = p_1$ and  $p_1$ is the first
Pontryagin class for the tangent bundle.  Here $c$ is the chiral central
charge -- the difference between the numbers of positive and negative
eigenvalues of the $K$-matrix.  In particular,
\begin{center}
\framebox{\parbox{.97\columnwidth}{if we choose $K$ in
\eqn{CSlatt} to be the $E_8$ matrix (\ie the integer matrix with even diagonal
and det$(K)=1$), then \eqn{CSlatt} realizes the 2+1D invertible topological
order with chiral central charge $c=8$.\cite{KW1458} }}\\
\end{center}
One may wonder, if the framing anomaly prevents us to have a lattice
realization of chiral CS theory with a non-zero central charge $c\neq 0$.  Our
construction shows that chiral $U(1)$ CS theory can always be realized on any
2+1D space-time lattice.  We think that this is possible because our space-time
lattice has an extra structure -- the branching
structure.\cite{C0527,CGL1314,CGL1204} It is possible that for the same
space-time lattice, if we choose different branching structures, the resulting
partition function $Z$ may be different.  This branching structure dependence
of partition function may represent the framing anomaly.

This research is
partially supported by NSF Grant No.  DMR-1506475 and DMS-1664412.

\vfill
\eject

\appendix

\section{Cochains and Cohomology}
\label{sec:CaC}

Let us first set some notation. We consider a three-dimensional simplicial
complex $M$, which we take to contain $0$-simplices (vertices), $1$-simplices
(links), $2$-simplices (faces), and $3$-simplices (faces). In this paper, we
consider our complex (also referred as space-time lattice in this paper) to
have matter fields living on the vertices, and gauge fields living on the
links. Moreover, we assume that these fields take values in an Abelian group. 

The matter fields, denoted $g_{i}$, form a map from the $0$-simplices of $M$ (or more formally, form a map from the space of $0$-chains of $M$) to the target space. This map is called a $0$-cochain, as it defines a linear map from the free abelian group on the $0$-simplices of $M$ to the target space. Similarly,  a gauge field $a_{ij}$ living on the links of the lattice defines a $1$-cochain. One can continue this, with $n$-cochains defining maps from the $n$-simplices to the target space. 

To see the explicit action of a $n$-cochain, let us label simplices by their vertices, so that an $n$-simplex is given by $[v_{0},..., v_{n}]$. Then an $n$-cochain $a$ assigns a target space element $a([v_{0}, ..., v_{n}])$  to any $n$-simplex. Furthermore, this map is multilinear over a formal sum of $n$-simplices with coefficients in $\mathbb{Z}$. For example, let $\sigma_{1} = [v_{0}, ..., v_{n}]$ and $\sigma_{2} = [w_{0}, ..., w_{n}]$. Then 
\begin{align}
a(2\sigma_{1} - \sigma_{2}) = 2a(\sigma_{1}) - a(\sigma_{2})
\end{align}

Given any $n$-cochain $a$, we can create an $n+1$ cochain $\dd a$ via:
\begin{align}
\dd a([v_{0}, ..., v_{n+1}]) = \sum_{i=0}^{n+1} (-1)^{i} a([v_{0}, ..., \hat{v_{i}}, ..., v_{n+1}])
\end{align}
where $\hat{v_{i}}$ means that we omit that index. One can then see that $\dd(\dd a)= 0$. This allows us to construct cohomology groups as the cohomology of the complex of $n$-chains:
\begin{align}
\leftarrow C^{n} \leftarrow C^{n-1} ...~C^{1}\leftarrow C^{0} \leftarrow 0
\end{align}
here the maps are just given by $d_{n}$, eg $d$ acting on elements of $C^{n}$, and the cohomology groups are just $\Im~d_{n}/\ker d_{n}$.

In the case that the target space is a ring, we have an additional structure called the cup product. The cup product takes an $m$-form $a_m$ and an $n$-form $b_n$ and returns an $n+m$ form $a\smile b$, defined by:
\begin{align}
&\ \ \ \ 
a_m\smile b_n([v_{0}, ..., v_{n+m}]) 
\nonumber\\
&= a_m([v_{0}, ..., v_{m}])b_n([v_{m}, ..., v_{m+n}])
\end{align}
Furthermore, when considered on cohomology classes, this cup product is graded
anticommutative. This means that there is a $n+m-1$ cochain $c_{m+n-1}$ such
that:
\begin{align}
a_m\smile b_n = (-1)^{nm}b_n\smile a_m + \dd c_{m+n-1}
\end{align}
To get an explicit expression for $c_{m+n-1}$, we need to introduce higher cup
product $a_m \hcup{k} b_n$ which gives rise to a $(m+n-k)$-cochain\cite{S4790}:
\begin{align}
\label{hcupdef}
&\ \ \ \
 a_m \hcup{k} b_n([0,1,\cdots,m+n-k])
\nonumber\\
&
 = 
\hskip -1em 
\sum_{0\leq i_0<\cdots< i_k \leq n+m-k} 
\hskip -3em  
(-)^p
a_m([0 \to i_0, i_1\to i_2, \cdots])\times
\nonumber\\
&
\ \ \ \ \ \ \ \ \ \
\ \ \ \ \ \ \ \ \ \
b_n([i_0\to i_1, i_2\to i_3, \cdots]),
\end{align} 
and $a_m \hcup{k} b_n =0$ for  $k<0$ or for $k>m \text{ or } n$.  Here $i\to j$
is the sequence $i,i+1,\cdots,j-1,j$, and $p$ is the number of permutations to
bring the sequence
\begin{align}
 0 \to i_0, i_1\to i_2, \cdots; i_0+1\to i_1-1, i_2+1\to i_3-1,\cdots
\end{align}
to the sequence
\begin{align}
 0 \to m+n-k.
\end{align}
For example
\begin{align}
&
 a_m \hcup1 b_n([0\to m+n-1]) 
 = \sum_{i=0}^{m-1} (-)^{(m-i)(n+1)}\times
\nonumber\\
&
a_m([0 \to i, i+n\to m+n-1])
b_n([i\to i+n]).
\end{align} 
We can see that $\hcup0 =\smile$.  
Unlike cup product at $k=0$, the higher cup product of two
cocycles may not be a cocycle. For cochains $a_m, b_n$, we have\cite{S4790}
\begin{align}
\label{cupkrel}
& \dd( a_m \hcup{k} b_n)=
\dd a_m \hcup{k} b_n +
(-)^{m} a_m \hcup{k} \dd b_n+
\\
& \ \ \
(-)^{m+n-k} a_m \hcup{k-1} b_n +
(-)^{mn+m+n} b_n \hcup{k-1} a_m 
\nonumber 
\end{align}
The above result also allows us to see that
the cup product interacts with $\dd$ in the familiar way:
\begin{align}
\dd(a_m\smile b_n) = (\dd a_m)\smile b_n + (-1)^{m}a_m\smile \dd b_n 
\end{align}
which we can interpret as the Leibniz rule. In the case that there is no boundary, we can interpret this as
yielding a form of integration by parts, so that $\dd a_m\smile b_n =
-(-1)^{m}a_m\smile \dd b_n $. We will abbreviate the cup product
with using `$\smile$,' so that $ab\equiv a\smile b$.

\section{Another form of $U(1)$ CS theory on lattice}

In the following, we are going to present another form of $U(1)$ CS theory on
lattice:
\begin{align}
\label{CSlatt1A}
Z=\int [\prod \dd a^{\RZ}_I]\ \Theta_{0}[a^{\RZ}_{I}]\Theta_{1}[a^{\RZ}_{I}]\Theta_{2}[a^{\RZ}_{I}]
\end{align}
where
\begin{align}
\Theta_{0}[a^{\RZ}_{I}] =& \ee^{\ii \pi K_{IJ} \int_{\cM^3} a^{\RZ}_{I} (\dd a^{\RZ}_{J} -\toZ{\dd a^{\RZ}_J})-\toZ{\dd a^{\RZ}_J}a^{\RZ}_I }
\\
\Theta_{1}[a^{\RZ}_{I}] =& \ee^{\ii \pi \sum_{I<J} K_{IJ} \int_{\cM^3} 
\toZ{\dd a^{\RZ}_I} \hcup{1}\toZ{\dd a^{\RZ}_J}
+ \toZ{\dd a^J}\hcup{2}\dd \toZ{\dd a^{\RZ}_I}
}
\nonumber\\
\Theta_{2}[a^{\RZ}_{I}]=&
\ee^{-\ii \pi K_{IJ} \int_{\cM^3}  a^{\RZ}_I\hcup{1}\dd \toZ{\dd a^{\RZ}_J}}
\ee^{- \int_{\cM^3} \frac{|\dd a^{\RZ}_I - \toZ{\dd a^{\RZ}_I}|^2}{g}} ,
\nonumber 
\end{align}
and $K_{II}=$ even integer, $K_{IJ}=$ integer.  We note that when $\dd
a^{\RZ}_I \approx 0$, \eqn{CSlatt1A} reduces to \eqn{eq:NaiveCS} (up to a
surface term).  The Maxwell term $\ee^{- \int_{\cM^3} \frac{|\dd a^{\RZ}_I -
\toZ{\dd a^{\RZ}_I}|^2}{g}}$ is included to make $\dd a^{\RZ}_I$ nearly an
integer if we choose $g$ to be small.  In the following, we will also assume
that $\cM^3$ has no boundary.  

To see that the path integral \eq{CSlatt1A} is invariant under \eq{gaugeZ}, we
proceed term-by-term. Under \eq{gaugeZ} the term $\Theta_{0}$ changes by a factor
\begin{align}
&\ \ \ \
\ee^{\ii \pi K_{IJ} \int_{\cM^3} n^{I} (\dd a^{\RZ}_{J} -\toZ{\dd a^{\RZ}_J})
-\toZ{\dd a^{\RZ}_J}n^I -(\dd n^J)a^{\RZ}_I -(\dd n^J)n^I }
\nonumber\\
&= \ee^{\ii \pi K_{IJ} \int_{\cM^3} n^{I} \toZ{\dd a^{\RZ}_J}
+\toZ{\dd a^{\RZ}_J}n^I +(\dd n^J)n^I }
\end{align}
Using \eqn{cupkrel}, we can rewrite the terms involving $\toZ{\dd a^{\RZ}_{I}}$:
\begin{align}
&\ \ \ \
\ee^{\ii \pi K_{IJ} \int_{\cM^3} n^{I} \toZ{\dd a^{\RZ}_J} +\toZ{\dd a^{\RZ}_J}n^I }
\nonumber\\
&= \ee^{\ii \pi K_{IJ} \int_{\cM^3} \dd n^{I} \hcup{1} \toZ{\dd a^{\RZ}_J} + n^{I} \hcup{1} \dd \toZ{\dd a^{\RZ}_J} }
\nonumber\\
&=\ee^{\ii \pi \sum_{I<J} K_{IJ} \int_{\cM^3} 
\dd n^{I} \hcup{1} \toZ{\dd a^{\RZ}_J} 
+\dd n^{J} \hcup{1} \toZ{\dd a^{\RZ}_I} 
}
\nonumber\\
&\ \ \ \ 
\ee^{\ii \pi  K_{IJ} \int_{\cM^3} 
n^I \hcup{1} \dd \toZ{\dd a^{\RZ}_J}
}
\nonumber\\
&=
\ee^{\ii \pi \sum_{I<J} K_{IJ} \int_{\cM^3} 
\dd n^{I} \hcup{1} \toZ{\dd a^{\RZ}_J} 
+
\toZ{\dd a^{\RZ}_I} \hcup{1} \dd n^{J} 
+
\dd n^{J} \hcup{2} \dd \toZ{\dd a^{\RZ}_I} 
}
\nonumber\\
&\ \ \ \ 
\ee^{\ii \pi K_{IJ} \int_{\cM^3} 
n^I \hcup{1} \dd \toZ{\dd a^{\RZ}_J}
}.
\end{align}
In the first equality, we used equation \eq{cupkrel} with $k=1$, while in the
third we used \eq{cupkrel} with $k=2$. We have also noted that since all
quantities in the exponential are integers times $\ii \pi$, we may dispense with
minus signs at well.  In the second equality, we have used the fact that the
diagonal elements of the $K$-matrix are even: $K_{II}=$ even.  Using the same
approach, we also have that:
\begin{align}
 \ee^{\ii \pi K_{IJ} \int_{\cM^3} (\dd n^J)n^I }
&=
\ee^{\ii \pi \sum_{I<J} K_{IJ} \int_{\cM^3} (\dd n^J)n^I +n^I\dd n^J }
\nonumber\\
&=
\ee^{\ii \pi \sum_{I<J} K_{IJ} \int_{\cM^3} \dd n^I\hcup{1} \dd n^J }.
\end{align}
Combining the above results, we see that $\Theta_{0}$ changes by:
\begin{align}
&\ \ \ \
\ee^{\ii \pi K_{IJ} \int_{\cM^3} n^{I} (\dd a^{\RZ}_{J} -\toZ{\dd a^{\RZ}_J})
-\toZ{\dd a^{\RZ}_J}n^I -(\dd n^J)a^{\RZ}_I -(\dd n^J)n^I }
\nonumber\\
&= 
\ee^{\ii \pi \sum_{I<J} K_{IJ} \int_{\cM^3} 
\dd n^{I} \hcup{1} \toZ{\dd a^{\RZ}_J} 
+
\toZ{\dd a^{\RZ}_I} \hcup{1} \dd n^{J} 
+
\dd n^{J} \hcup{2} \dd \toZ{\dd a^{\RZ}_I} 
}
\nonumber\\
&\ \ \ \ 
\ee^{\ii \pi \sum_{I<J} K_{IJ} \int_{\cM^3} \dd n^I\hcup{1} \dd n^J }
\ee^{\ii \pi K_{IJ} \int_{\cM^3} 
n^I \hcup{1} \dd \toZ{\dd a^{\RZ}_J}
}
\end{align}
Similarly, under the gauge transformation \eq{gaugeZ}, the term $\ee^{\ii \pi \sum_{I<J} K_{IJ} \int_{\cM^3}
\toZ{\dd a^{\RZ}_I} \hcup{1}\toZ{\dd a^{\RZ}_J} } $ changes by a factor
\begin{align}
 \ee^{\ii \pi \sum_{I<J} K_{IJ} \int_{\cM^3} 
\dd n^{I} \hcup{1} \toZ{\dd a^{\RZ}_J} 
+
\toZ{\dd a^{\RZ}_I} \hcup{1} \dd n^{J} 
+
\dd n^I\hcup{1} \dd n^J 
}
\end{align}
while the term $ \ee^{\ii \pi \sum_{I<J} K_{IJ} \int_{\cM^3} \toZ{\dd
a^{\RZ}_J}\hcup{2}\dd \toZ{\dd a^{\RZ}_I} } $ changes by:
\begin{align}
 \ee^{\ii \pi \sum_{I<J} K_{IJ} \int_{\cM^3} 
\dd n^{J} \hcup{2} \dd \toZ{\dd a^{\RZ}_I} 
}
\end{align}
Hence $\Theta_{1}$ changes by:
\begin{align}
 \ee^{\ii \pi \sum_{I<J} K_{IJ} \int_{\cM^3} 
\dd n^{I} \hcup{1} \toZ{\dd a^{\RZ}_J} 
+
\toZ{\dd a^{\RZ}_I} \hcup{1} \dd n^{J} 
+
\dd n^I\hcup{1} \dd n^J 
}\\
 \ee^{\ii \pi \sum_{I<J} K_{IJ} \int_{\cM^3} 
\dd n^{J} \hcup{2} \dd \toZ{\dd a^{\RZ}_I} 
}
\end{align}
Finally, $\Theta_{2}$ changes by:
\begin{align}
 \ee^{-\ii \pi K_{IJ} \int_{\cM^3} 
n^I \hcup{1} \dd \toZ{\dd a^{\RZ}_J}
}
\end{align}
and so the product $\Theta_{0}\Theta_{1}\Theta_{2}$ is indeed invariant under $\eq{gaugeZ}$, but only when $\cM^3$ has no boundary.

\bibliography{../../bib/all,../../bib/publst}

\end{document}